\documentclass[11pt]{article}

\usepackage[utf8]{inputenc} 
\usepackage[T1]{fontenc}    
\usepackage{hyperref}       
\usepackage{url}            
\usepackage{booktabs}       
\usepackage{amsfonts}       
\usepackage{nicefrac}       
\usepackage{microtype}      
\usepackage{graphicx}
\usepackage{natbib}
\usepackage{doi}
\usepackage{csquotes}
\usepackage{siunitx}

\newcommand{\Description}[2][]{%
}

\title{Reducing the rate of personal insults in social media with bystander bots}

\author{%
  Libby Hemphill\\
  University of Michigan\\
  \texttt{libbyh@umich.edu}
  \and
  Lingyao Li\\
  University of South Florida\\
  \texttt{lingyaol@usf.edu}
  \and
  Ryan Burton\\
  University of Michigan
  \and
  David Jurgens\\
  University of Michigan\\
  \texttt{jurgens@umich.edu}
}

\date{}

\begin{document}
\maketitle

\begin{abstract}
Prompted by previous research on strategies for reducing interpersonal conflict and addressing problematic behaviors in online communities, a randomized controlled trial on Reddit compared various responses for reducing the rate of personal insults users post to the site. We generated replies from five deescalation strategies and used an automated procedure for posting them as replies to insulting comments. The findings reveal that automated replies to insults can effectively reduce their rate. Appreciation performed best. Not all strategies performed well, though. We conclude that automated responses are a viable tool for addressing some problematic behaviors. We discuss their potential utility and limitations.
\end{abstract}

\section*{Introduction}
Hate speech, harassment, and other anti-social behaviors that make social media conversations toxic are pervasive, can discourage people from participating in online conversation, and therefore limit the positive impacts social media can have. Online communities struggle to live up to their democratic potential~\cite{Papacharissi2004-zu} in part because some users actively work to poison the conversation so that others cannot engage~\cite{Sood2012-jt}. Both "anti-social" users and ordinary users can poison online conversations ~\cite{Cheng2017-kk}. Such users employ a variety of tactics to poison conversations, including trolling, harassing, insulting, threatening, intimidating, berating, and defaming other users. Together, these tactics effectively discourage participation and jeopardize healthy discussions. They also disproportionately influence vulnerable communities and members of historically marginalized groups~\cite{Mossie2020-uo}. 

These negative social impacts demand efforts to address these behaviors. Platforms that host online communities, such as Facebook, Reddit, and Twitter, have developed strategies, often called ``content moderation''~\cite{Roberts2019-qb}, to address these attempts to poison conversations. Retributive responses, such as removing content and sanctioning users, are common strategies for moderating content ~\cite{Pater2016-xv}. However, the fact that these toxic behaviors continue tells us that current approaches may not work. 
Retributive moderation strategies can even cause some unintended consequences for those users who are marginalized and reinforce a sense that social media platforms are places where users could be targeted for their speech, beliefs, or identity \cite{Myers_West2018-cm}. In addition, the variety of toxic behaviors suggests that a one-size-fits-all mitigation approach will not work---trolling and harassment have different underlying motivations, for instance, so an effective treatment for one will not likely work for the other. 

What then, instead of takedowns and sanctions, could reduce the rate of toxic behaviors in online communities? Bystander intervention is effective at addressing incidents of conflict offline, and bystander intervention in cyberbullying contexts is receiving increasing research attention \cite{Taylor2019-nx,Kazerooni2018-tf}. Bullies often use insults as part of their attempts to exert dominance and exclude the bullied from their social networks. Inactive bystanders can increase the deleterious impacts of bullying on a victim~\cite{Brody2021-ep}, but bystanders who intervene on behalf of victims effectively reduce bullying's impacts and occurrence~\cite{Polanin2012-rt}.

Our project examines whether a bystander intervention can work in an online community. We focus specifically on personal insults because they functionally socially to denigrate and exclude \cite{Barber2016-jl}. We used a bot to respond to insults with a comment that employed one of several specific anti-exclusion strategies. We chose personal insults because they are markers of contempt, an emotion whose social function is exclusion ~\cite{Fischer2007-xw}. By belittling others, insulters show contempt for their targets. Targets of contempt tend to withdraw from social groups \cite{Mackie2000-ep} and to dissolve partnerships \cite{Gottman1994-mw}. Because contempt produces these behaviors in its targets, it functions to socially exclude individuals from communities \cite{Hutcherson2011-hm}. 

We based our experiment plan on the Bystander Intervention Model (BIM) \cite{Latane1970-xj} which identifies five stages: notice, appraise, take responsibility, choose response, and intervene. We attempted to automate each stage of the BIM with an ensemble of tools (i.e., insult detection models, automated interventions). In this paper, we focus on the intervention step and its impacts.


Exclusion is the opposite of what we want in a healthy online community. Social exclusion negatively affects those excluded and can lead to less prosocial behavior \cite{Twenge2007-jv} and even self-defeating behavior \cite{Twenge2002-yk}. Feeling excluded also inhibits our individual ability to self-regulate \cite{Baumeister2005-ko} and can produce a spiral of aggressive retaliation \cite{Twenge2001-rs}. Interrupting the spiral could improve online discussion dramatically; therefore, we experimented with strategies to reduce insults and the social exclusion they generate.

We designed an experiment on Reddit to determine whether bystander interventions reduce personal insults. The general design of our experiment consists of three key steps: (1) detecting insults in a comment thread, (2) posting a reply designed to de-escalate the conversation, and (3) comparing the rate of insults before and after the intervention. We investigated two specific research questions:
\begin{enumerate}
    \item Do users who receive bot replies in response to an insult they posted post fewer insults in the future?
    \item Are the bystander bots' intervention strategies equally effective?
\end{enumerate}


\section*{Related Work}

\subsection*{Detection and Impacts of Toxicity}
Identifying and addressing toxic content are critical aspects of managing social media platforms. One area of research has focused on examining the negative impacts of toxic messages on communities and society. For instance, Mohan et al. \cite{Mohan2017-sn} revealed a high negative correlation between community health and online toxicity in a study of Reddit conversations. Hateful speech has severe consequences for historically marginalized groups, discouraging them from participating in online conversation. For example, \cite{Keighley2022-ir} surveyed 175 individuals from the LGBTQ+ community and found that online hate speech could cause significant emotional damage to LGBTQ+ young people, resulting in reports of sadness, shame, and feelings of inferiority.

Many prior studies have focused on developing models to automatically identify and flag toxic messages ~\cite{Cheng2017-kk, Georgakopoulos2018-yh,Rupapara2021-rm, Fortuna2020-td}. Initially, researchers developed lexicons or dictionaries for such purposes \cite{Gitari2015-cy,Wiegand2018-se}. This method involves compiling a list of terms and phrases, such as the Hurtlex, a multilingual lexicon of hate words \cite{Tontodimamma2023-xb}. Once the lexicons were developed, rule-based methods, such as sentence-level subjectivity detection \cite{Gitari2015-cy} were used to filter social media posts containing items from these lexicons. However, rule-based systems struggle to capture messages' context or intent and to recognize forms of harmful content that are not included in the collected lexicons \cite{Dixon2018-mf}. 

More recent studies have leveraged machine learning classifiers with natural language processing (NLP) techniques for the detection of hate speech and toxicity \cite{Zhang2024-eo, Li2024-cj, Rupapara2021-rm, Taleb2022-aj, Vaidya2020-ki, Juuti2020-gy, Nobata2016-dp, Davidson2017-te}. For example, Rupapa et al. \cite{Rupapara2021-rm} leveraged support vector machine and logistic regression with Term Frequency-Inverse Document Frequency (TF-IDF) for feature extraction to detect toxic comments on social media. Their model achieved an accuracy of 0.94 based on annotated Wikipedia comments. Taleb et al. \cite{Taleb2022-aj} experimented with multiple deep learning classifiers (e.g., CNN, LSTM, GRU) with word embeddings techniques (e.g., GloVe, FastText), and their results revealed that these models could reach an accuracy of 0.94 and a recall of 0.93 and 0.94 for non-toxic and toxic respectively, based on the same annotated Wikipedia dataset. Nearly all of these studies focus solely on detection and do not discuss what should happen in communities \textit{after} toxic behavior.

\subsection*{Content Moderation and Bystander Intervention}

The negative social impacts of toxicity demand efforts to address toxic behaviors on social media. Social media platforms have developed strategies, often called ``content moderation''~\cite{Roberts2019-qb}. Content moderation involves a range of techniques and approaches to regulate user-generated content and ensure that it adheres to community guidelines and standards. The most common approaches to responding to bad behavior are to remove content and punish users \cite{Atreja2023-rl, Lo2020-nw}. Removing content involves deleting comments or other forms of user-generated content that violate community guidelines; punishing users can involve a range of sanctions from temporary suspensions to permanent bans. However, such content moderation can have unintended consequences, particularly for those who are already marginalized in society. A prior study revealed that content moderation practices, such as flagging and banning, could reinforce a perception among users that platforms are places where they could be targeted for their speech, beliefs, or identity \cite{Myers_West2018-cm}. This perception can be ultimately detrimental to users' experiences and make it challenging for platforms to manage toxic content at scale. Therefore, addressing alternative moderation strategies beyond flagging and removal are necessary. 

Bystander intervention has been shown to be a useful strategy to mitigate harmful behaviors on social media \cite{Munger2017-ji, Fischer2011-nb}. Bystander intervention refers to the active involvement of individuals who witness a harmful or problematic situation and take steps to address it \cite{Fischer2011-nb}. Research on bystander interventions has explored various topics, including evaluating the effectiveness of bystander interventions in different circumstances~\cite{Edwards2019-ho, Polanin2012-rt}, identifying promoters and barriers to intervention~\cite{Bennett2014-it}, and examining predictors and factors associated with bystander interventions~\cite{Lindegaard2022-xb,Henson2020-mm}. In particular, previous research has illustrated that bystanders can effectively reduce racist slurs on Twitter~\cite{Munger2017-ji}, sexual violence~\cite{Mujal2021-iz}, and other emergencies~\cite{Fischer2011-nb}. 

Latane and Darley \cite{Latane1970-xj} pointed out that when there are more bystanders, each individual feels less responsible for responding. In large online communities, diffusion of responsibility likely inhibits bystander interventions. Recent work on bystander interventions in cyberbullying explained why bystanders select aggressive (e.g., saying mean things about the bully) or constructive responses (e.g., comforting the victim) ~\cite{Moxey2020-sd} and that bystanders are more likely to respond when many bullies are involved~\cite{Kazerooni2018-tf}.

In summary, research on toxicity online has focused on detection and rarely addressed prevention or intervention. Bystander intervention has showed promise in social conflicts offline and in reducing racism online. We address these prior literatures by deploying a detection model to identify personal insults, a specific type of toxic behavior, and using a bystander bot to respond with messages designed to reduce insults.

\section*{Methods}
\subsection*{General Experiment Design}
As presented in Figure 1, the general experiment design includes three key steps: (1) use a narrowly specified personal insult detection model to identify insults in a comment thread, (2) use the bystander bot to post a reply to deescalate the conversation, and (3) collect user comments before and after the intervention and measure the rate of insults.

\begin{figure}[h]
    \centering
    \includegraphics[width=0.75\textwidth]{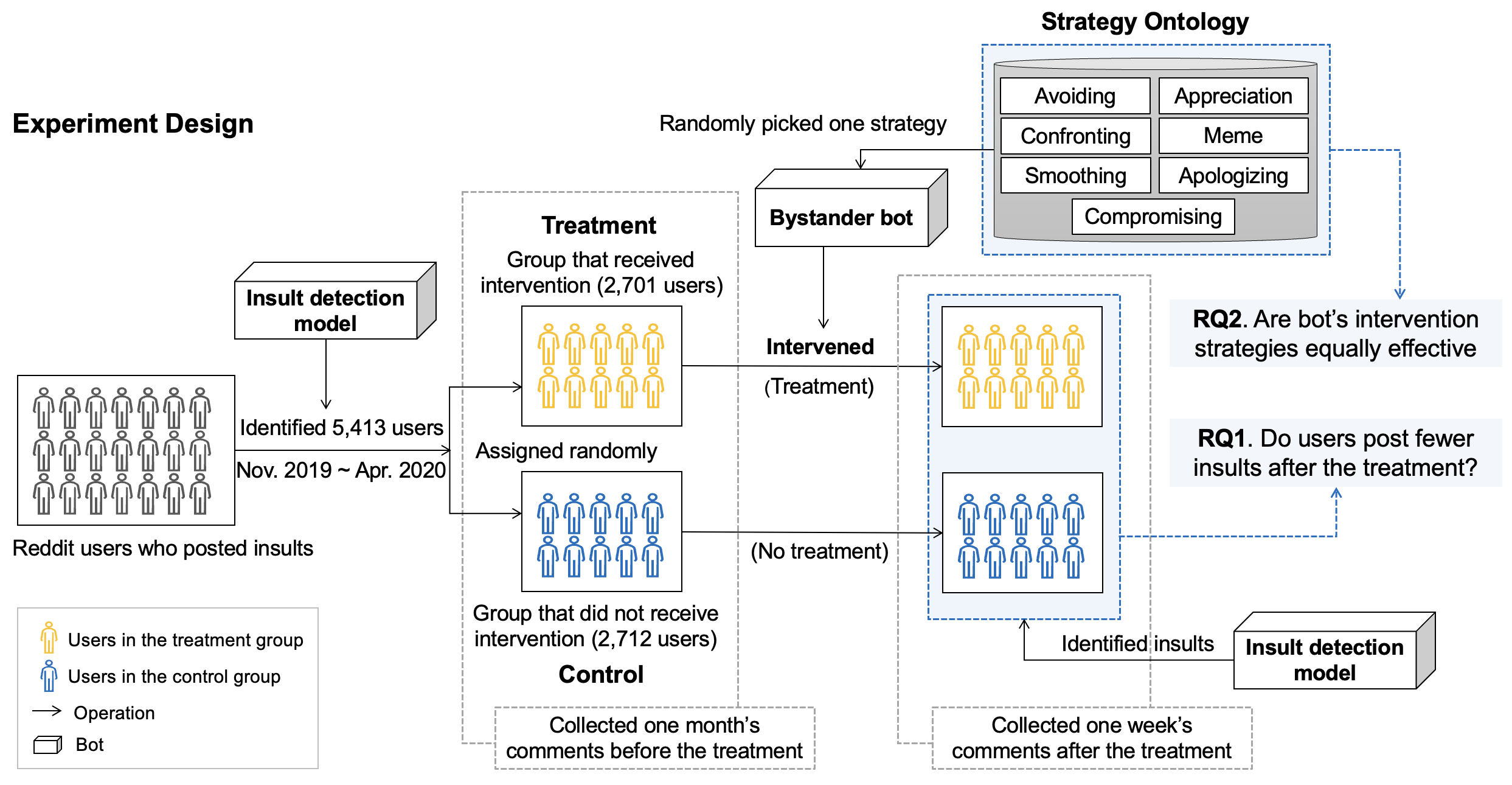}
    \caption{Overview of our experimental design}
    \label{fig:design}
    \Description[Overview of our experimental design]{A detailed flowchart of our experimental design and process}
\end{figure}

We monitored comments using the Python Reddit API Wrapper (PRAW)
\footnote{URL: \url{https://praw.readthedocs.io/en/latest}}
We consumed all new comments on Reddit using 'r/all' and passed them to our insult detection model. We randomly assigned users who posted insults to either the treatment or control group the first time our system detected an insult in comments they authored. Then, when a comment was flagged as an insult by the insult detection model, we checked (1) whether the user who posted the insult was already on our treatment or control list, (2) whether the subreddit was on our "ok-to-post" list, and (3) whether our bot had already posted in the same thread.

Our “ok-to-post” list included all subreddits with at least 10,000 subscribers that did not explicitly forbid bots; that allowed low-karma users to post; that were not age-restricted; that were not quarantined; that were not in our ``do-not-post`` topic list. Before automating our experiment, we ran a manual pilot study in which a team member manually posted a candidate reply to insults our model identified. The team member spent hours each day reading Reddit and monitored community responses to our replies. The goals of the pilot were to test the insult detection model and to make sure we didn't cause immediate and obvious harm. During our pilot study, we noticed a baseline level of toxicity and patterns of acceptable teasing that suggested bystander bots would not be welcome in subreddits about mental health, sports, gaming, or men’s rights. If the insult was a candidate for response (i.e., the user was in the treatment group, the subreddit was in our “ok-to-post” list, and our bot had not posted yet), we randomly selected one of 35 messages to post in response.

We used multiple Reddit accounts to post our automated messages. We created Reddit accounts during the pilot period. Each of these accounts was given a random username (i.e., a username suggested by Reddit when we created each account) and was subscribed to multiple different subreddits. Before the experiment, our team members used these accounts to comment on posts on Reddit to generate a history of activity. Team members continued to post from the accounts throughout the experiment period; the volume of comments posted manually by the team was higher for all accounts than the volume of bot-posted comments. When manually commenting, we aimed to post only on threads where insults had not occurred and to use inoffensive comments to generate account history. Some examples of manual posts include “Support local businesses! We can do it.” in a subreddit about people helping one another, and “I have the Pro 12.9, and it is heavy. The 2nd Gen Apple Pencil makes a big difference though, so I’d get the smaller Pro and the better Pencil since your whole point is note taking.” in a subreddit for general conversation.

When users were assigned to the treatment group, we used our bystander bot to post the automated messages from its strategy list (Figure 1). The bystander bot randomly selected a response from a set of 35 options and posted it as a reply to the insulting comment. All of our responses were written by the research team, and they each employed one of the seven de-escalation strategies defined below.

After the intervention, we continued to monitor all Reddit comments and labeled them as "insult" or "not insult" throughout the experiment period using the same insult detection model. This process generated a dataset of all Reddit comments, their content, karma, user metadata, and insult classification. From that dataset, we also generated the following measures for result analysis: (1) count of each user’s comments, (2) count of each user’s comments that were insults, and (3) percentage of each user’s comments that were insults.

\subsection*{Detecting Insults} 
\subsubsection*{Why build a new model?}
Machine learning and AI models that detect ``bad'' behavior online are widely researched and available, so why did we make our own model? We built a custom model with new training data because existing models are either unavailable or do not perform well on the personal insult detection task. Specifically, they generate many false positives where the detection model identifies insult where human annotators do not. General ``toxicity'' models identify myriad types of comments that can impact online discussions \cite{Fortuna2020-td,Waseem2017-ch}. We wanted to limit the scope of our intervention to focus on a single type of toxic content (i.e., a single detection ``subtask'' in \cite{Waseem2017-ch}'s language) to reduce our chances for errors and increase the likelihood that we could observe an effect if it exists. To create a model with sufficient precision, we needed an insult-specific model rather than a generic toxicity or harassment model. Two other insult detection models were not available \cite{Sood2012-jt} or included generic ``negative comments'' in their insult category\cite{Goyal2022-qd}. Existing training data presented similar challenges because their labels contain overlapping categories, are not specific to our subtask of personal insult detection, or exhibit unacceptable levels of interannotator variation \cite{Goyal2022-qd,Vidgen2019-ia}.

Therefore, to detect insults among the stream of
Reddit comments being posted in almost real time, we employed a classifier
built using a combination of rule-based and machine learning. A
comment was considered to be an insult if and only if our machine learning
model classified it as such, and if it also contained a \emph{you construction},
that is, a slur directed at the second person, as determined by our
rule-based classifier. As such,

\[
    [\![ Classify_{ML}(Comment_i) = insult ]\!] \wedge [\![ YouConstruction \in Comment_i \Rightarrow 1 ]\!] \Rightarrow insult
\]

\subsubsection*{Rule-Based Classification: ``You Construction''}
The \textit{you construction} rule requires that (a) a second-person pronoun be present and (b) likely in the subjective case for a comment to be classified as a personal insult (e.g., ``you were late" contains a \textit{you construction} but ``that response was yours" does not). We included this rule to improve the precision of our model.

This rule was developed by parsing Part-of-Speech (POS) tags and sentence dependency. Using these two techniques, we first tokenized a comment based on POS tags and identified the pronouns and verbs. Then we located the elements of siblings, children, and grandchildren based on sentence dependency and considered these elements as “direct relationships” of the identified pronouns or verbs. Last, we classified a comment as an insult if the direction relationships of pronouns or verbs contained any insulting words from our predefined list\footnote{We sourced insult words from \url{https://github.com/melsherief/hate_speech_icwsm18/blob/master/hate_keywords.txt} and \url{http://www.bannedwordlist.com/lists/swearWords.txt}. We manually reviewed these lists and kept words where multiple authors reached consensus.}.

\subsubsection*{Machine Learning Classification}
For machine learning classification, we used a fine-tuned BERT model
\cite{Devlin2018-qy}. At the time of the experiment, BERT provided
state-of-the art performance for natural language classification.
This experiment required binary classification; a label for each comment
was intended to be in the set $\left\{ insult, \neg insult \right\}$.

We labelled our training and test set with crowdsourced workers via
Amazon Mechanical Turk (MTurk). We collected 4547 comments from Reddit to label, of which 10 comments were required by each worker. HITs were approved when the labels of a worker had an agreement of at least 75\%.

The instructions provided to Turkers defined what was expected of their
labeling as a \textit{personal insult}. This description was as follows:

\blockquote{
    \begin{enumerate}
        \item A comment contains a personal insult if an attack is directed against a person rather than the position they are maintaining.
        \item Insults include hate speech. Hate speech is speech that attacks a person on the basis of gender, race, religion, sexual orientation, or disability. In addition to hate speech, personal insults include \textit{ad hominem} attacks and disrespect towards an individual.
        \item Personal insults can be direct/explicit (e.g., “You are pathetic.”) or indirect/implied (e.g., “Of course you would politicize a hurricane. Pathetic!”).
        \item Mark the comment as a personal insult even if it is a mild personal attack.
        \item Note that the presence of profanity does not necessarily mean that a comment is an insult.
    \end{enumerate}
}

We assigned each comment to three workers. Whenever a worker finished annotating, we approved and saved it along with the referenced comment IDs. If one worker failed annotation (e.g., did not complete annotation in time), we assigned the comment to another randomly selected worker. Once all three annotations were completed, we finalized the label of this comment following the majority of these three annotations. As a result, we annotated 4,000 comments for the training dataset and 547 comments for the testing dataset. 

\subsubsection*{Model Performance}
We provide comparisons of various insult detection models on our test data (N = 547, of which 272 are inults). For purposes of our experiment, a low number of false positives is especially important; the bot should not reply unless an insult occurs. 

As a result, the fine-tuned BERT model achieved a precision of 0.70 and 0.88, a recall of 0.92 and 0.60, and an F1-score of 0.80 and 0.72 for non-insult and insult classes, respectively. To validate the performance of the BERT model, we also tested other baseline ML models, including Na\"ive Bayes and support vector machine in conjunction with Term-Frequency Inverse Document Frequency (TD-IDF). The fine-tuned BERT model produced more balanced F1 scores for both classes and a better overall testing accuracy than these two baseline models. Due to its superior performance, we applied the fine-tuned BERT model to detect Reddit insults. We acknowledge that our model's ability to precisely and reliably detect insults is an important factor in our experiment's success. Our model isn't perfect, and therefore it sometimes indicated we should respond when we should not and missed opportunities for the bot to intervene.

\subsection*{Posting a Reply as a Bystander Bot}
When users were assigned to the treatment group, our bystander bot randomly selected a response from a set of 35 options\footnote{A complete list of responses and their categories is available in the supplemental material.} and posted it as a reply to the insulting comment. Our research team manually generated the reply options. Each reply used one of the seven de-escalation strategies defined below. We solicited multiple rounds of feedback from colleagues and summer research assistants not involved in the project to refine our list of candidate replies. Each reply is designed to use a de-escalation strategy from psychology literature and to sound like a Reddit comment. 

\begin{itemize}
    \item \textit{Apologize} responses either apologize on behalf of one of the arguers or apologize for the tense situation the arguers find themselves in. These responses include ``I’m sorry this conversation got so dark, but please try to keep it more civil" and ``I’m sorry someone is getting under your skin, but it won’t help to keep arguing like this."
    \item \textit{Appreciate} responses emphasize a positive aspect of the argument such as its passion or its interesting subject matter, while still urging a return to civility. These include “We appreciate your opinions, but not your aggressive language” and “Thanks for your willingness to discuss opposing viewpoints, but I’d appreciate it if you kept it friendly.”
    \item \textit{Avoid} responses either offer outside commentary on the argument or urge others to stay out of it, often expressing disappointment or disinterest in disputes. These responses include “this conversation has taken an unfortunate turn,” “move it along folks, nothing to see here,” or “this is too much, abandon thread.”
    \item \textit{Compromise} responses are geared towards finding a middle ground between multiple people arguing, as well as offering a solution or resolution to the argument. Our compromising responses end with “let’s try to stay calm,” “can’t we agree to disagree?” or “please keep it civil.”
    \item \textit{Confront} responses are slightly more aggressive and scolding towards the arguing commenters. Unlike compromising responses, this tactic uses more direct and authoritative language in making resolution suggestions such as “end this conversation” or “stop fighting in here.”
    \item \textit{Meme} responses are simply GIFs or images from popular culture that emphasize respect or staying calm. Most of these memes are sarcastic and comedic ways of intervening in an argument by lightening a situation or poking fun at its ridiculousness.
    \item \textit{Smooth} responses are the most polite out of all our intervention tactics, offering support and rationality by using empathetic statements and positive language. They include “Hey man, do you need a hug?” or “It seems like we’re all worked up. Deep breath everybody.”
\end{itemize}

Blake and Mouton \cite{Blake1964-jk} proposed these strategies (except \textit{meme}) in their work on conflict management strategies in organizations. Related work recommends similar approaches to de-escalate interpersonal conflict in intimate relationships \cite{Gottman1994-mw}. Gottman also recommends \textit{appreciation} as a way to promote admiration and respect and provide an antidote to contempt in relationships \cite{Gottman1999-fj}. We added \textit{meme} because they appear often on Reddit as meta-commentaries on ongoing conversation, and we wanted to test the impact of humor and online-specific communication. 

In our experiment, all users who posted insults received no response (control group) or a reply designed to change their behavior (treatment group). None of bystander bots’ responses are “neutral” or “innocuous” -- ignoring insulting behavior implicitly condones it. This means that there is no true control present in our study design. However, we are still able to detect the influence of our responses by comparing them to one another: the variation of effects between the responses suggests that the responses themselves, and not their presence, are responsible for the observed effects. Users received only one treatment from the bystander bot. Each Reddit post received only one treatment, regardless of the number of comments containing insults that appeared in comments responding to it.

\section*{Results}
Our experiment, outlined in Figure \ref{fig:design}, started by applying insult detection models and identified 26,298 Reddit users who posted personal insults in our ``ok-to-post`` subreddits between November 2019 and April 2020. We randomly assigned users to the treatment group (n = 2955). The users in the treatment group received a single, random response from our bystander bots. We collected comments from all users in both groups one month before and one week after their identified insult and labeled all comments using the same insult detection model. Here, we focus on changes in user behavior in the week immediately after intervention.

\begin{table}[h]
  \caption{Number of users who received each intervention strategy and the maximum and mean number of insults they posted in the week following intervention.} 
  \label{tab:counts} 
    \begin{tabular}{lrrr}
        \hline
        Intervention strategy & N & Max & Mean \\
        \hline
        Control & 21048 & 192 & 1.14 \\
        Apologize & 258 & 20 & 1.20 \\
        Appreciate & 269 & 31 & 1.24 \\
        Avoid & 447 & 50 & 1.24 \\
        Compromise & 474 & 43 & 1.18 \\
        Confront & 509 & 32 & 1.28 \\
        Meme & 479 & 43 & 1.17 \\
        Smooth & 272 & 36 & 1.25 \\
        \hline
    \end{tabular}
\end{table}

These sub-samples vary because the concepts themselves exhibit more variation, and we used more comments of some types than others. Our goals were to identify a general effect for bystander bots, if available, and to learn more about the potential for individual strategies and replies. We were not aiming, given the size of our sample and the variation we expected, to formally test hypotheses about individual strategies. We were also not aiming to develop a state-of-the-art insult detection model. Rather, we needed a model that reliably identified insults with a low false positive rate; the purpose of the insult detection model was to find good candidates for replies and not to outperform other models in identifying bad behaviors.

\subsection*{Community Response}

\begin{figure}[h]
    \centering
    \includegraphics[width=0.75\textwidth]{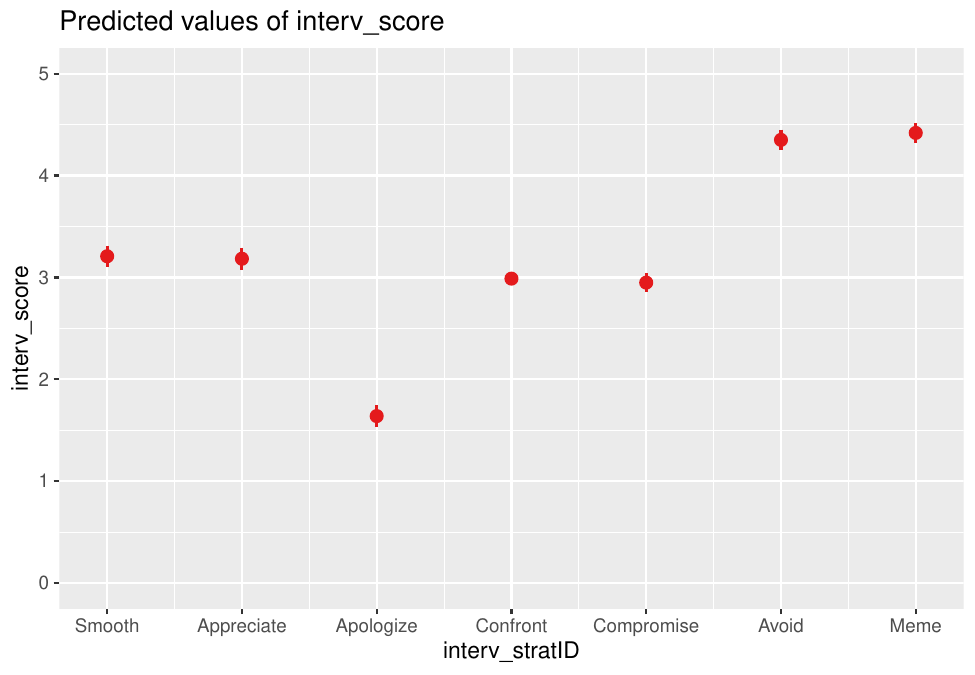}
    \caption{Visualizing the impact of intervention strategy on the karma a bystander bot's comment received using a marginal effects plot.}
    \label{fig:marginal-effects}
    \Description{Visualizing the impact of intervention strategy on the karma a bystander bot's comment received using a marginal effects plot.}
\end{figure}

In general, our interventions received positive feedback from the Reddit community (see Table \ref{tab:karma}. We measured this by modeling the relationship between `karma`, the ratio of up-votes to down-votes a comment receives, and the intervention strategies. In these models, we controlled for subreddits. 

\begin{table}[!htbp] \centering 
  \caption{Predicting comment karma using intervention strategies. The regression model was \textit{karma $=$ intervention\_strategy $+$ subreddit}. We omit the subreddit coefficients for brevity.} 
  \label{tab:karma} 
\begin{tabular}{@{\extracolsep{5pt}}lc} 
\\[-1.8ex]\hline 
\hline \\[-1.8ex] 
 & \multicolumn{1}{c}{\textit{Dependent variable:}} \\ 
\cline{2-2} 
\\[-1.8ex] & karma \\ 
\hline \\[-1.8ex] 
 Appreciate & 1.545$^{***}$ \\ 
  & (0.056) \\ 
 Avoid & 2.711$^{***}$ \\ 
  & (0.053) \\ 
 Compromise & 1.310$^{***}$ \\ 
  & (0.049) \\ 
 Confront & 1.349$^{***}$ \\ 
  & (0.050) \\ 
 Meme & 2.780$^{***}$ \\ 
  & (0.051) \\ 
 Smooth & 1.569$^{***}$ \\ 
  & (0.055) \\ 
 Constant & $-$1.711 \\ 
  & (1.235) \\ 
\hline \\[-1.8ex] 
Observations & 293,192 \\ 
R$^{2}$ & 0.645 \\ 
Adjusted R$^{2}$ & 0.645 \\ 
Residual Std. Error & 4.274 (df = 292602) \\ 
F Statistic & 903.479$^{***}$ (df = 589; 292602) \\ 
\hline 
\hline \\[-1.8ex] 
\textit{Note:}  & \multicolumn{1}{r}{$^{*}$p$<$0.1; $^{**}$p$<$0.05; $^{***}$p$<$0.01} \\ 
\end{tabular} 
\end{table}

\subsection*{Treatment Effects}
We evaluated the effectiveness of the intervention by examining 
\begin{enumerate}
    \item whether treatment resulted in users posting fewer total insults
    \item whether it reduced the proportion of insults users posted.
\end{enumerate}

We analyzed both count and percentage because we wanted to understand both the absolute and relative impact of our intervention. For instance, if users posted fewer insults but also posted fewer comments (lower count but same percentage) or the same number of insults but fewer comments (same count but higher percentage), our intervention would suppress commenting but not insulting. Similarly, studying the percentage of insults naturally controls for comment frequency. We used hurdle regression \cite{Cragg1971-hy} to model first, whether individuals posted insults after intervention and second, the probability of more insults after. In the subsections that follow, we address our research questions about bystander bots in general, provide descriptive statistics, and discuss the potential of individual intervention strategies.

\subsubsection*{Bystander Bots Do Not Impact Likelihood of Posting an Insult}
We first calculated a logistic regression to determine whether control or treatment users differed in their likelihood of posting an insult at all (in the next week). We found that, holding the previous month's insult rate constant, the groups did not differ in their insult likelihoods (OR = 1.013).

\subsubsection*{Bystander Bots Do Not Reduce the Number of Insults Treated Users Post}
We calculated a negative binomial generalized linear model (NB-GLM) to explore the question of whether insulters posted fewer insults (by count) after bot interventions. Independent variables in this model include the experiment condition (i.e., whether a user received a treatment), the percentage of insults in the previous month, and loadings from a principal component analysis (PCA) based on users' prior participation in subreddits. The PCA used two variables: users and the subreddits in which they posted before the experiment. Low-dimensional representations of the communities in which users are active are known to capture behavioral regularity \cite{Waller2021-wx}. As communities have different levels of toxicity \cite{Park2021-qw}, we include these factors to capture relative differences in baseline toxicity in a user's communities.  NB-GLM results indicated main effects for both treatment (OR = $1.064^{***}$) and the previous month's insult rate (OR = $1.079^{***}$). We excluded the PCA loadings from the table for brevity.

The interaction between the treatment group and the previous month's rate of insults (OR = $0.987^{***}$) indicates that the users who received a bot response had a smaller increase in the insult rate compared to those in the control group.

\begin{table}[!htbp] \centering
\caption{Negative binomial generalized linear model predicting count of insults in the week following detection; PCA loadings are excluded for brevity. Table reports odds ratios.}
\label{tab:nb-glm-count}
\begin{tabular}{@{\extracolsep{5pt}}lc}
\\[-1.8ex]\hline
\hline \\[-1.8ex]
& \multicolumn{1}{c}{\textit{Dependent variable:}} \\
\cline{2-2}
\\[-1.8ex] & Next week insult count \\
\hline \\[-1.8ex]
Treatment & 1.064$^{***}$ \\
& \\
Previous month insult percentage & 1.079$^{***}$ \\
& \\
Treatment*Previous month insult percentage & 0.987$^{***}$ \\
& \\
\hline \\[-1.8ex]
Observations & 9,229 \\
Log Likelihood & $-$83,332.250 \\
$\theta$ & 2.226$^{***}$ (0.024) \\
Akaike Inf. Crit. & 166,882.500 \\
\hline
\hline \\[-1.8ex]
\textit{Note:} & \multicolumn{1}{r}{$^{*}$p$<$0.1; $^{**}$p$<$0.05; $^{***}$p$<$0.01} \\
\end{tabular}
\end{table}

\begin{figure}[h]
    \centering
    \includegraphics[width=0.75\textwidth]{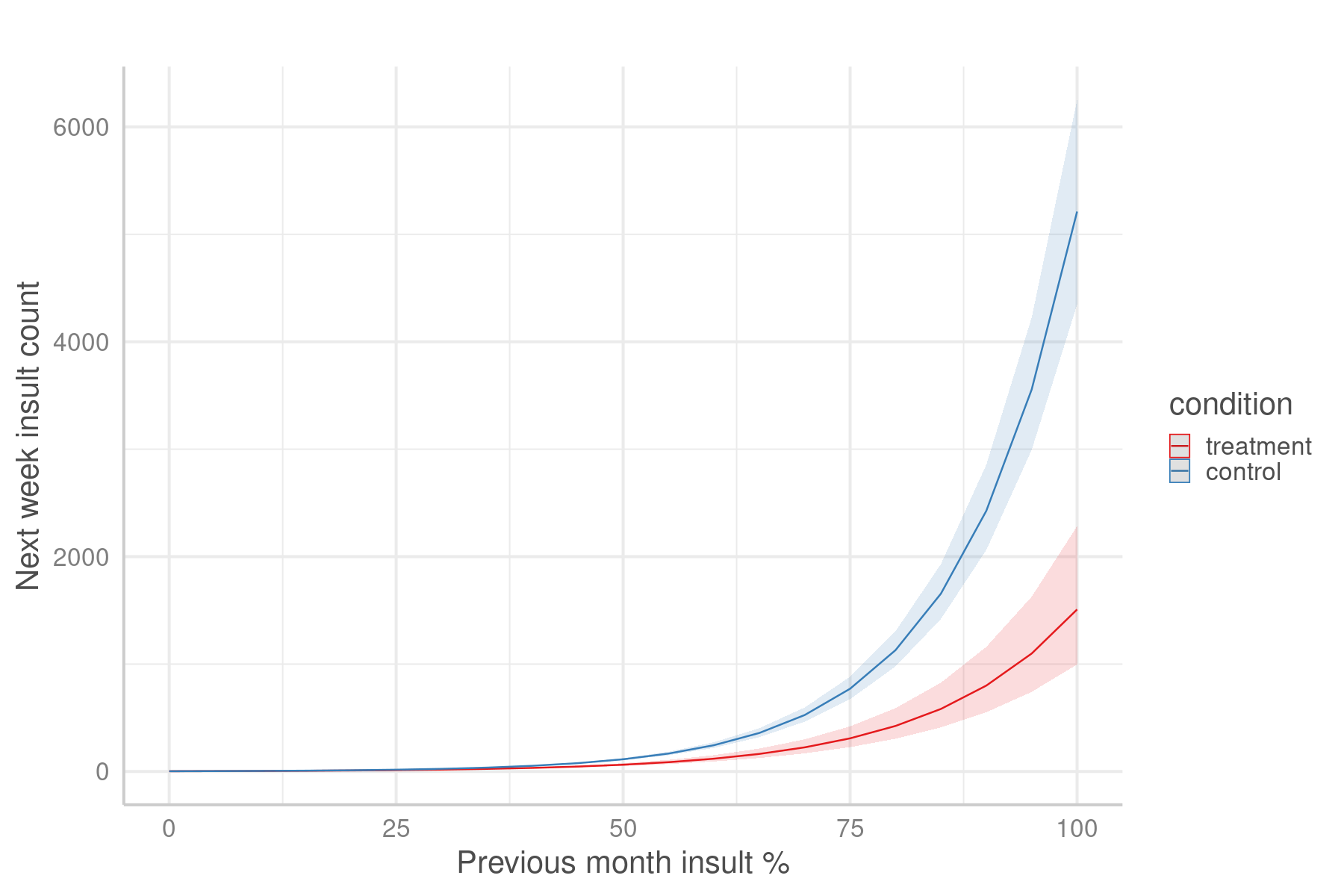}
    \caption{Predicted insults by count after personal insult detection}
    \label{fig:count}
    \Description[Predicted insults by count after personal insult detection]{A plot of predicted insult count where y-axis represents next week insult count and x-axis represent previous week's insult count; the control condition shows a steeper slope than the treatment condition.}
\end{figure}

Figure \ref{fig:count} makes this difference more apparent. Bystander interventions had positive effects, but only among users who posted a high percentage of insults before treatment.

\subsubsection*{Bystander Bots Do Not Impact the Rate at Which Users Post Insults}
Our earlier model shows that bystander bot interventions can reduce insults by count for users who post a high percentage of insults. To ensure that we were not suppressing participation generally, we needed to measure impacts on both insults and innocuous comments; examining rates lets us do just that.

\begin{figure}[h]
    \centering
    \includegraphics[width=0.75\textwidth]{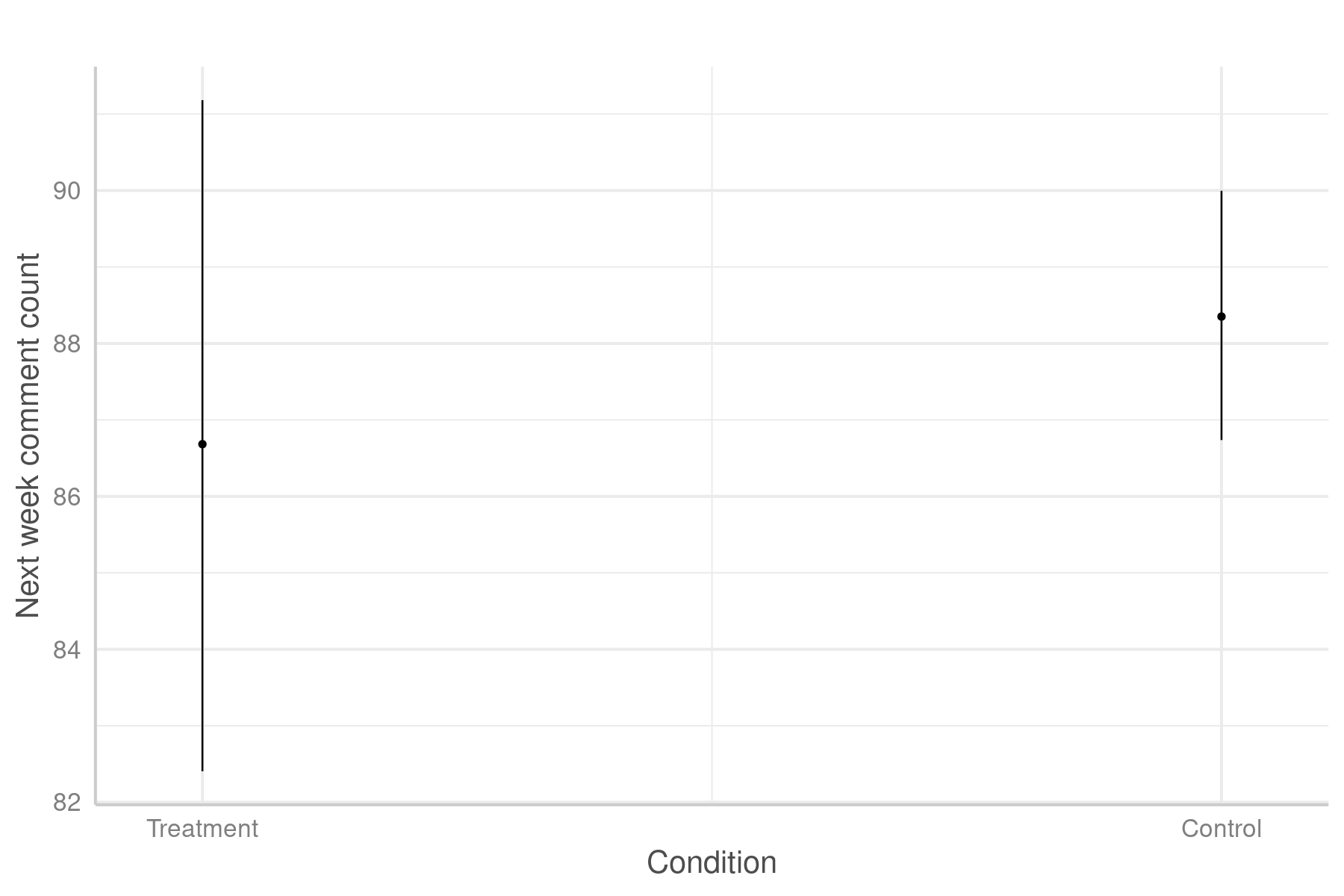}
    \caption{Total comments in the week following detection}
    \label{fig:comment_count}
\end{figure}

We did not observe a difference in the number of comments posted between the treatment and control groups -- the bystander messages did not impact overall comment activity (see Figure \ref{fig:comment_count}. The independent variables are the same as those of the NB-GLM, and the dependent variable is the percentage of insults after treatment. Since the dependent variable is bounded and continuous (i.e., between 0 and 1), we built a robust linear model (RLM) rather than a GLM to test the hypothesis. The model outputs are presented in Table \ref{tab:rlm}.

\begin{table}[!htbp] \centering
  \caption{Predicting the percentage of insults after intervention using RLM}
  \label{tab:rlm}
 \begin{tabular}{@{\extracolsep{5pt}}l S[table-align-text-post=false,table-format=1.4]}

\\[-1.8ex]\hline
\hline \\[-1.8ex]
 & \multicolumn{1}{c}{\textit{Dependent variable:}} \\
\cline{2-2}
\\[-1.8ex] & {Next week insult \%} \\
\hline \\[-1.8ex]
 Treatment& 0.008 \\
 Previous month insult \% & 0.251$^{***}$ \\
 Treatment:Previous month insult \% & 0.028$^{***}$ \\
 Constant & 0.413$^{***}$ \\
\hline \\[-1.8ex]
Observations & {23,756} \\
Residual Std. Error & {0.932 (df = 23752)} \\
\hline
\hline \\[-1.8ex]
\textit{Note:}  & \multicolumn{1}{r}{$^{*}$p$<$0.1; $^{**}$p$<$0.05; $^{***}$p$<$0.01} \\
\end{tabular}
\end{table}


\subsubsection*{Bystander Bot Strategies Exhibit Variance in their Impacts}
Our results so far indicate that bystander bots can be effective interventions, but only for some users; for most users they have no significant impact. We applied seven strategies of bystander interventions to address personal insults: avoid, compromise, confront, smooth, apologize, appreciate, and meme. The variance among our bots' strategies means that we can begin to examine the differential impacts of different strategies.

First, we analyzed whether individual bystander bot strategies influenced whether treated users posted insults at all (see Table \ref{tab:likelihood-by-strat}). The results of our logistic regression appear in Table \ref{tab:insulted-after-strat}. We found that, when we controlled for prior behavior, no strategy significantly reduced the likelihood of a future insult. \textit{Appreciate} bots actually increased the likelihood of a future insult.

\begin{table}[!htbp] \centering
  \caption{Predicting the likelihood of posting an insult at all after specific intervention strategies (odds ratios)}
  \label{tab:insulted-after-strat}
\begin{tabular}{@{\extracolsep{5pt}}l S[table-align-text-post=false,table-format=1.4] }
\\
\hline
\hline \\
 & \multicolumn{1}{c}{\textit{Dependent variable:}} \\
\cline{2-2}
& {Posted an insult in the next week} \\
\hline \\
 Apologize & 0.835 \\
 Appreciate & 1.270$^{*}$ \\
 Avoid & 1.136 \\
 Compromise & 0.894 \\
 Confront & 1.147 \\
 Meme & 1.068 \\
 Smooth & 1.003 \\
 Previous month insult \% & 1.103$^{***}$ \\
 Apologize:Previous month insult \% & 1.115$^{**}$ \\
 Appreciate:Previous month insult \% & 0.945\\
 Avoid:Previous month insult \% & 1.003 \\
 Compromise:Previous month insult \% & 1.126$^{***}$ \\
 Confront:Previous month insult \% & 1.037 \\
 Meme:Previous month insult \% & 0.968 \\
 Smooth:Previous month insult \% & 0.971 \\

\hline \\[-1.8ex]
Log Likelihood & {$-$14,672.770~~~} \\
Akaike Inf. Crit. & {~~~~29,587.540} \\
\hline
\hline \\[-1.8ex]
\textit{Note:}  & \multicolumn{1}{r}{$^{**}$p$<$0.05; $^{***}$p$<$0.01} \\
\label{tab:likelihood-by-strat}
\end{tabular}
\end{table}


\subsubsection*{Appreciate and Confront Interventions Likely Reduce the Number of Insults Users Post}
We calculated an NB-GLM model to determine whether strategies had differential impacts on those who did post insults. In addition to the independent variables for each intervention strategies, we included the percentage of insults in the prior month and its interaction with each strategy as model inputs. The dependent variable was the number of insults a user posted in the following week. The results are presented in Table \ref{tab:nb-glm-strat}.

Users who received \textit{Meme} responses posted more insults after; users who received \textit{Appreciate} or \textit{Confront} responses posted fewer. We observed no significant differences for the other strategies.


\begin{table}[!htbp] \centering
  \caption{Predicting the count of insults after receiving individual conflict resolution strategies using NB-GLM (odds ratios)}
  \label{tab:nb-glm-strat}
  \begin{tabular}{@{\extracolsep{5pt}}l S[table-align-text-post=false,table-format=1.4] }

\\[-1.8ex]\hline
\hline \\[-1.8ex]
 & \multicolumn{1}{c}{\textit{Dependent variable:}} \\
\cline{2-2}
\\[-1.8ex] & {Next week insult count} \\
\hline \\[-1.8ex]
 Apologize & 0.984 \\
 Appreciate & 1.318$^{***}$ \\
 Avoid & 0.972 \\
 Compromise & 0.879$^{***}$ \\
 Confront & 1.096$^{***}$ \\
 Meme & 0.991 \\
 Smooth & 1.325$^{***}$ \\
 Previous month insult \% & 1.079$^{***}$ \\
 Apologize:Previous month insult \% & 1.002 \\
 Appreciate:Previous month insult \% & 0.949$^{***}$ \\
 Avoid:Previous month insult \% & 0.997 \\
 Compromise:Previous month insult \% & 1.005 \\
 Confront:Previous month insult \% & 0.966$^{***}$ \\
 Meme:Previous month insult \% & 1.020$^{***}$ \\
 Smooth:Previous month insult \% & 1.000 \\
  \hline \\[-1.8ex]
Log Likelihood & {$-$83,271.400~~} \\
Akaike Inf. Crit. & {166,784.800} \\
\hline
\hline \\[-1.8ex]
\textit{Note:}  & \multicolumn{1}{r}{$^{*}$p$<$0.1; $^{**}$p$<$0.05; $^{***}$p$<$0.01} \\
\end{tabular}
\end{table}


\subsubsection*{Avoid, Meme, Smooth, and Appreciate Interventions May Reduce the Rate of Insults Users Post}

\begin{figure}[h]
    \centering
    \includegraphics[width=0.75\textwidth]{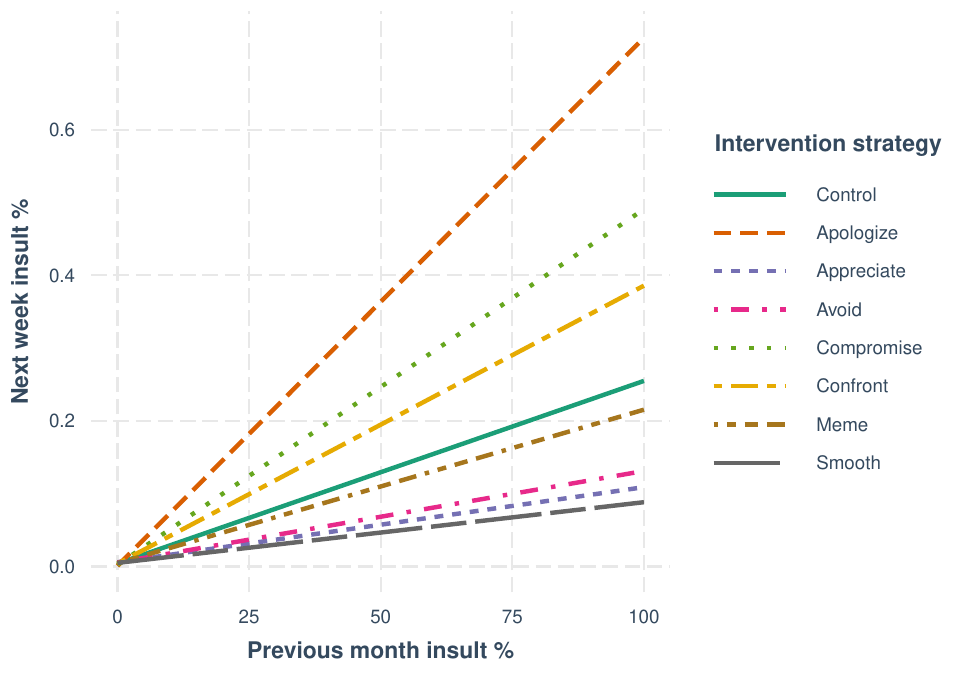}
    \caption{Predicted insults by percentage after the experiment with different strategies.}
    \label{fig:pred-pct-strat}
\end{figure}

We calculated a RLM model to investigate whether these intervention strategies could equally reduce proportion of insults after the treatment. The independent variables are the same as in the NB-GLM, while the dependent variable is the percentage of insults after the experiment. The results are presented in Table \ref{tab:rlm-strat} and Figure \ref{fig:pred-pct-strat}, respectively.

Users who received \textit{Avoid}, \textit{Meme}, \textit{Smooth}, or \textit{Appreciate} interventions posted lower proportions of insults in the week after the intervention. Users who received the other strategies posted higher proportions of insults. These effects were stronger among users with higher proportions of insults before the experiment. See Table \ref{tab:rlm-strat} for specifics.

Only \textit{Appreciation} could help reduce insults by both measures ($p<0.05$ in both cases). \textit{Apologize} consistently showed an adverse effect.

\begin{table}[!htbp] \centering
  \caption{Predicting the percentage of insults after specific intervention strategies using RLM}
  \label{tab:rlm-strat}
\begin{tabular}{@{\extracolsep{5pt}}lc} 
\\[-1.8ex]\hline 
\hline \\[-1.8ex] 
 & \multicolumn{1}{c}{\textit{Dependent variable:}} \\ 
\cline{2-2} 
\\[-1.8ex] & next\_week\_insult\_prop \\ 
\hline \\[-1.8ex] 
Apologize & $-$0.003$^{***}$ \\ 
  & (0.001) \\ 
Appreciate & 0.002$^{**}$ \\ 
  & (0.001) \\ 
Avoid & 0.002$^{**}$ \\ 
  & (0.001) \\ 
Compromise & $-$0.002$^{**}$ \\ 
  & (0.001) \\ 
Confront & $-$0.0001 \\ 
  & (0.001) \\ 
Meme & 0.0004 \\ 
  & (0.001) \\ 
Smooth & 0.001 \\ 
  & (0.001) \\ 
 prev\_month\_insult\_pct & 0.003$^{***}$ \\ 
  & (0.00002) \\ 
Apologize:prev\_month\_insult\_pct & 0.005$^{***}$ \\ 
  & (0.0002) \\ 
Appreciate:prev\_month\_insult\_pct & $-$0.001$^{***}$ \\ 
  & (0.0001) \\ 
Avoid:prev\_month\_insult\_pct & $-$0.001$^{***}$ \\ 
  & (0.0002) \\ 
Compromise:prev\_month\_insult\_pct & 0.002$^{***}$ \\ 
  & (0.0002) \\ 
Confront:prev\_month\_insult\_pct & 0.001$^{***}$ \\ 
  & (0.0001) \\ 
Meme:prev\_month\_insult\_pct & $-$0.0004$^{***}$ \\ 
  & (0.0001) \\ 
Smooth:prev\_month\_insult\_pct & $-$0.002$^{***}$ \\ 
  & (0.0002) \\ 
 Constant & 0.004$^{***}$ \\ 
  & (0.0001) \\ 
\hline \\[-1.8ex] 
Observations & 23,756 \\ 
Residual Std. Error & 0.009 (df = 23740) \\ 
\hline 
\hline \\[-1.8ex] 
\textit{Note:}  & \multicolumn{1}{r}{$^{*}$p$<$0.1; $^{**}$p$<$0.05; $^{***}$p$<$0.01} \\ 
\end{tabular} 
\end{table}

\section*{Discussion}


The prevalence of harmful behavior, such as personal insults, on social media platforms overwhelms the human moderators who address it. We drew on previous research highlighting the importance of bystander intervention \cite{Polanin2012-rt,Bennett2014-it,Kazerooni2018-tf} and strategies to manage conflict \cite{Blake1964-jk, Gottman1994-mw} offline to design bystander bots to help reduce personal insults on Reddit and used a field experiment to test their efficacy. Our findings have three main takeaways: bystander bots did not cause clear backlash effects; appreciation is an effective strategy for bystander bots to employ; and issuing apologies on behalf of someone else is not effective.

Our analysis reveals that bystander bots do not cause a backlash effect (i.e., users do not respond to the intervention with increased vitriol). Specifically, we found that users engaged in less aggressive or insulting behavior the following week after bot intervention than their prior behavior would predict. Among frequent insulters, bystander posts were effective at reducing the \textit{count} of insults but not their \textit{rate} which could mean that the insulters became less active after treatment but not less anti-social. These findings are consistent with previous research suggesting that moderation bots can help reduce the rate of abusive posts on Reddit \cite{Young2018-wr}, possibly because bots can help create a sense of accountability and normative influence among users. When users are aware that their insults are being monitored and that there are consequences for displaying aggressive behavior (e.g., that someone will respond to their insult), they may be more likely to conform to social norms, avoid engaging in such behavior, or become less active in spaces where norms are enforced.

Our findings also imply that bystander bots' intervention strategies play different roles in reducing insults. Among the seven proposed strategies, \textit{Appreciate} is particularly effective in reducing the frequency and rate of insults. \textit{Appreciate} involves recognizing positive behaviors and actions while still emphasizing the importance of respect in communication. By doing so, a more supportive and inclusive environment can be created, encouraging users to interact positively. Appreciation can help build relationships and foster a sense of community rather than isolating those who post insults. Existing research on racism in the workplace \cite{Okun2001-ps}, organizational development \cite{Leiter2012-gz, Osatuke2009-xu}, and marital conflict conflict management \cite{Gottman1999-fj} identify appreciation as a tool for addressing conflict in a variety of settings. Our results indicate that appreciation can also be an effective strategy among strangers in public discussions.

\textit{Meme}, \textit{Avoid}, and \textit{Smooth} strategies also showed promise reducing insults among frequent insulters. Memes were designed to be humorous interventions that lowered the temperature in discussions. Our pilot and experience using Reddit suggested that memes were a common tool Redditors use to diffuse tensions or change the subject. Avoidance, where the bot signaled its intent to leave a conversation because of insults, was likely effective because it explicitly acknowledge the negative impacts insults had on other participants. Smoothing strategies were the most gentle intervention types; their impacts suggest that subtle nudges toward better behavior may be effective.

\textit{Apologize} and \textit{compromise} were not effective when employed by our bystander bots. Apologies signal efforts to restore social order and to acknowledge when someone has violated a norm \cite{Tavuchis1991-se}. It may be that only first-person apologies (where we apologize for own actions) are effective for reconciliation. Apologizing on behalf of someone else does not require the person who violated the norm to do any work to repair a social relationship.

Similarly, \textit{compromise} may not adequately address the emotional harm insults cause and therefore fails to restore social harmony. Compromising is a specific approach to conflict resolution in which parties try to reach a middle ground that satisfies the conflicting parties \cite{Lippitt1982-wm}. Successful compromise requires some cooperation from each party in the conflict, and it may be that insulters and targets were not willing to cooperate, especially when the compromise is offered by an outsider.

Both strategies fail to hold users accountable for their insulting behaviors or to promote genuine changes because they do not require insulters to acknowledge that they violated a norm or are in conflict with others. The target of the insult may not feel heard or validated, and the person who posted the insult may not learn from their mistake. Bot apologies and attempts at compromise can even exacerbate the situation by creating additional conflict or resentment towards the bot.

\subsection*{Implications for Content Moderators}

Our findings have significant implications for how moderators can design and leverage bystander bots to mitigate insulting behaviors in online communities. First, bot moderation has been shown to be useful in online communities \cite{He2021-lx, Kiene2020-wi}. We suggest moderators consider using bystander bots to quickly identify and address insulting comments. Bystander bots are efficient, scalable, and consistent. \textit{Appreciate} was the most effective bystander bot intervention strategy. While other strategies such \textit{Avoid}, \textit{Confront}, \textit{Meme}, or \textit{Smooth} may work in some situations, their efficacy may not be consistent.

We recommend that moderators avoid using \textit{apologize} or \textit{compromise} strategies when using bystander bots. These strategies could potentially escalate the situation by creating additional tension. Human moderation is still necessary to ensure that users are held accountable for their behavior \cite{Kim2021-cu}. Bots can serve as a useful tool in this process but should be used in conjunction with other measures, such as clear community guidelines and user education on appropriate behavior \cite{Jhaver2019-we}.

\subsection*{Limitations}

Several important limitations could affect the generalizability and validity of our findings. First, our insult detection model affected which comments we detected and whether we replied. A different detection model may have produced different results because it would select different moments for the bystander bot to intervene. Each decision point in our model training could impact its performance -- our training and testing data, the embedding and transformers, our "you construction" rule. MTurkers may not be the most sensitive or precise annotators for personal insults, and they were our primary source of annotations for training. While BERT has shown promise in previous research in text classification \cite{Devlin2018-qy,Garrido-Merchan2023-ie}, its performance can be affected by limited training data, overfitting, and limited generalization. Our model detected only a fraction of all insults posted on Reddit and therefore limits the generalizability of our findings. For instance, insults that do not contain a second-person pronoun were not detected, and responding to such implied insults may require different strategies to be effective.

Our bot responses also impacted our experiment's results. We used manually-authored replies and randomly-selected them at response time. We were unable to choose our responses based on the type of insult, insulter, target, or even subreddit where the insult occurred. More specific bystander bots that are tuned to the unique norms of the community in which they are deployed may fare differently. Because we used random rather than strategic responses, our experiment likely underestimates how effective bots can be at addressing insults.

The time frame of the data we analyzed also impacts our findings. We provide insights into the effectiveness of intervention strategies within this time window (one week after intervention). We controlled for the number of comments over all and the rate of comments in our regression models, but other changes in user behavior like taking a Reddit break or a sudden increase in their commenting around a specific event could impact that week's posts. The size of our sample likely produced enough between-user variance to account for some of the impacts of time, but global events could have impacted what users were talking about and the tone they took with one another. Furthermore, we did not examine the potential impact of these strategies on participants' attitudes over time, as all observations made in this paper were based on one week after intervention. Therefore, future research could expand the analysis window beyond a week to investigate the potential decay of the effects of these strategies.

Finally, we used only one platform, Reddit, during our experiment. Reddit users are known to be predominantly young, male, and digital in their news preferences \cite{Barthel2016-jj}. This demographic bias could affect the prevalence and nature of insults and responsiveness to bystanders on the platform. Our findings may not generalize to other online communities with different user demographics or social norms.

\subsection*{Future Work}
Ongoing and future research could build on our findings by addressing the limitations mentioned above. For example, a logical next step is to replicate our approach with a different detection model. A more accurate model may limit the chance that the bot replies when it should not (i.e., our bot may have responded erroneously). A second project could experiment with a more sophisticated response selection approach. Instead of randomly selecting replies, the system could choose a reply based on the features of the conversation in which it would intervene. In addition, future research could explore the sustainability and long-term impact of different intervention strategies on reducing insulting behavior on Reddit. It would also be valuable to examine the generalizability of this research by applying bystander bots to intervene on other social media platforms.


\section*{Acknowledgments}
The work was supported in part by a Mozilla Research Grant.

\bibliographystyle{unsrtnat}
\bibliography{paperpile}

\appendix

\begin{table}[]
\centering
\caption{The list of comments from intervention bots}
\label{tab:comments-intervention bots}
    \begin{tabular}{l p{.8\linewidth}} 
    \\[-1.8ex]\hline 
        \hline\\[-1.8ex] 
        Strategy & Intervention comment\\ 
        \hline
    Compromise & I agree with you on this stuff but you're being too aggressive. Please try to be less hostile. \\
    Compromise & I get what you're saying. Let's try to stay calm though.\\
    Compromise & Let's just try to calm down and be a little more level-headed here. \\
    Compromise & Guys, it's fine for us to have differences of opinion, but try to keep it respectful. \\
    Compromise & It's fine for us to have differences in opinion, but lets try to keep it respectful. \\
    Avoid & No one is here for the bashing...including me. \\
    Avoid & Yikes, this is too much. Watch out people. \\
    Avoid & I think it would be best if we all just leave you two here to keep bickering. \\
    Apologize & I'm sorry this conversation is so distressing to you, but let's try to be decent to each other and not just get into arguments. \\
    Apologize & I know what it's like to have someone get under your skin, it really sucks, and I'm sorry it's happening to you right now. What you have to know is that it doesn't help to keep arguing with that person though. \\
    Apologize & :( I'm sorry this conversation seems to have made your day a little harder, but by continuing to be rude you're just furthering the cycle of unpleasantness. \\
    Appreciate & I love this sub when we stay friendly. \\
    Smooth & It's usually not this nasty in here, right? We're nice? \\
    Smooth & Whoo, this got intense! Everyone alright? \\
    Appreciate & You are in the right here, but using insults is not going to change anyone else's opinion. Thanks! \\
    Smooth & Please cool it. This back and forth isn't productive for anyone. \\
    Confront & What's with the insults? There's no need for that in here. \\
    Confront & Stop fighting in here y'all. it ruins the thread. \\
    Confront & Give each other some respect or end this convo. \\
    Confront & Why so many insults? \\
    Confront & Stop trying to prolong this fight. it’s pointless. \\
    Confront & Just stop with the insults already. \\
    Confront & Slow down there speed racer.  \\
    Confront & Time for an internet break here. \\
    Confront & Hold your horses cowboy, keep your insults in your pocket. \\
    Confront & Lmao this is ridiculous.  \\
    Confront & Could we just...not do this?  \\
    \hline 
    \hline \\[-1.8ex] 
    \end{tabular} 
\end{table} 

\begin{table}
    \begin{tabular}{ll}
    \hline
    \includegraphics[width=15em]{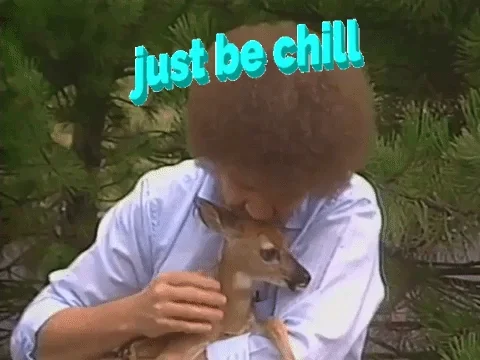} & \includegraphics[width=15em]{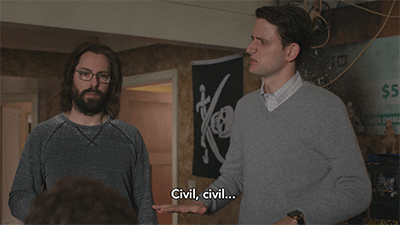}  \\ 
    \includegraphics[width=15em]{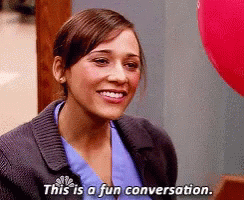} & \includegraphics[width=15em]{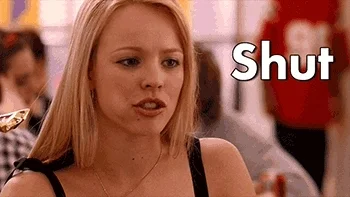} \\ 
    \includegraphics[width=15em]{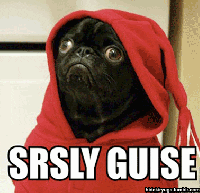} & \includegraphics[width=15em]{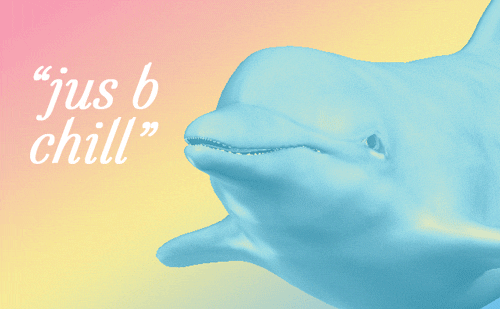} \\
    \includegraphics[width=15em]{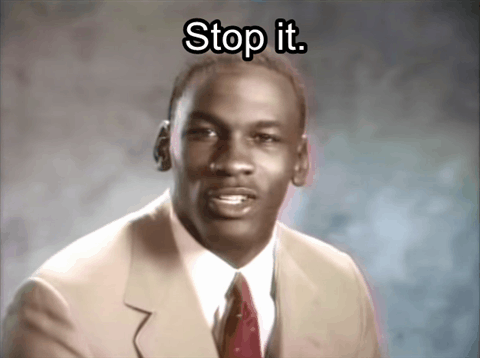}& \includegraphics[width=15em]{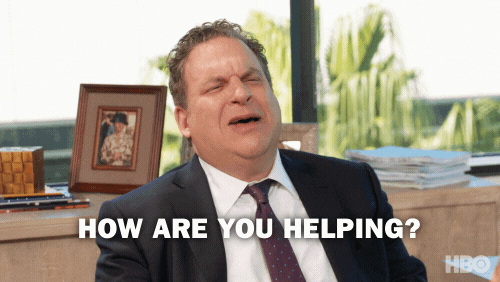} \\
    \hline
    \end{tabular}
\caption{Images of the bot's meme strategies. Note: some images were animated GIFs, and we show only one frame here.}
\end{table} 
    
\end{document}